\documentclass[aps,prd,twocolumn,superscriptaddress,nofootinbib,10pt]{revtex4-1}   

\usepackage[utf8]{inputenc}
\usepackage{amssymb}
\usepackage{amsmath}
\usepackage{amsfonts}
\usepackage{graphicx}
\usepackage{color}
\usepackage{xspace}
\usepackage{comment}
\usepackage{hyperref}
\usepackage[normalem]{ulem}

\usepackage[section]{placeins}
\usepackage{afterpage}

\usepackage{float}
\usepackage{slashed}
\usepackage{ulem}
\usepackage{appendix}

\usepackage{cancel}

\usepackage{multirow,rotating}
\usepackage[dvipsnames]{xcolor}
\usepackage{orcidlink}   

\begin{document}
	
      \title{Updated sensitivities to heavy neutral leptons at the LHC far detectors and SHiP}

	   \author{Zeren Simon Wang\,\orcidlink{0000-0002-1483-6314}}
	   \email{wzs@hfut.edu.cn}
      \affiliation{School of Physics, Hefei University of Technology, Hefei 230601, People’s Republic of China}

      \author{Yu Zhang\,\orcidlink{0000-0001-9415-8252}}
      \email{dayu@hfut.edu.cn}
      \affiliation{School of Physics, Hefei University of Technology, Hefei 230601, People’s Republic of China}

	\begin{abstract}       
        In recent years, a number of experiments dedicated to searches for long-lived particles (LLPs) have been proposed, approved, or have entered operation. While the sensitivities of these experiments to various LLP scenarios have been extensively studied, key aspects--such as detector geometries, background estimates, and projected operational durations--for several facilities, including \texttt{MATHUSLA}, \texttt{ANUBIS}, and \texttt{SHiP}, have undergone significant updates. In this work, we implement the latest experimental designs in the Displaced Decay Counter tool for calculating detector acceptances and signal-event yields, and re-evaluate their sensitivity reach to one of the most widely studied LLP scenarios, namely minimal heavy neutral leptons.
 	\end{abstract}

 	\keywords{}


	\maketitle
    \noindent

\noindent \textbf{Introduction.} The Large Hadron Collider (\texttt{LHC}) at CERN culminated in the discovery of a Standard-Model (SM)-like Higgs boson in 2012~\cite{ATLAS:2012yve,CMS:2012qbp}.
Continuing searches for physics beyond the SM (BSM) have so far yielded no positive signals, while increasingly stringent lower bounds have been placed on the masses of heavy fundamental particles predicted in various BSM scenarios, such as supersymmetry (SUSY); see Refs.~\cite{Nilles:1983ge,Martin:1997ns} for reviews, and also Refs.~\cite{Kwon:2025fgb,Sonneveld:2025ckk,Baer:2025zqt} for recent summaries of the current status and future prospects of SUSY searches at the \texttt{LHC} and high-luminosity \texttt{LHC} (\texttt{HL-LHC}).
In particular, current lower bounds on superpartner masses at the TeV scale imply a large degree of fine-tuning, rendering supersymmetry no longer a natural solution to the hierarchy problem of the SM~\cite{Giudice:2017pzm}.

This situation has led both theorists and experimentalist to re-examine existing search strategies for BSM physics.
One possibility is that new physics does not manifest itself through heavy states that decay promptly after production, but instead through long-lived particles (LLPs).
LLPs are predicted to give rise to a rich variety of exotic and unconventional signatures at terrestrial experiments, including colliders and beam-dump facilities; see Refs.~\cite{Alimena:2019zri,Lee:2018pag,Curtin:2018mvb,Beacham:2019nyx,Jeanty:2025wai,Alimena:2025kjv} for reviews on both theoretical and experimental perspectives.

The \texttt{LHC} collaborations have undertaken extensive efforts to search for LLPs; see Refs.~\cite{ATLAS:2025lfx,ATLAS:2025pak,CMS:2025urb,CMS:2025qkk,LHCb:2021dyu,LHCb:2025ymr} for a selection of recent results reported by \texttt{ATLAS}, \texttt{CMS}, and \texttt{LHCb}.
Although no LLP predicted in BSM scenarios has been discovered to date, increasingly strong exclusion limits have been obtained.
In addition to the \texttt{LHC} main detectors, several dedicated experiments for LLP searches have been proposed, approved, or have already begun operation.
In these experiments, LLPs produced at the interaction point (IP) are expected to travel macroscopic distances before potentially decaying inside dedicated fiducial volumes.
Such experiments include a class of far detectors installed at distances of roughly 10--600~m from different \texttt{LHC} IPs, as well as a beam-dump experiment, the Search for Hidden Particles (\texttt{SHiP})~\cite{SHiP:2015vad,Alekhin:2015byh,SHiP:2018xqw,SHiP:2021nfo,Albanese:2878604,SHiP:2018yqc,SHiP:2025ows}.

Among the proposed \texttt{LHC} far-detector experiments, \texttt{FASER}~\cite{Feng:2017uoz,FASER:2018eoc,FASER:2022hcn} has been operating since 2022 and has already published its first results~\cite{FASER:2023tle,FASER:2024bbl}.
Other proposals include \texttt{ANUBIS}~\cite{Bauer:2019vqk,ANUBIS:2025sgg}, \texttt{CODEX-b}~\cite{Gligorov:2017nwh,Aielli:2019ivi}, \texttt{FACET}~\cite{Cerci:2021nlb}, \texttt{FASER2}~\cite{Salin:2927003}, \texttt{MoEDAL-MAPP1} and \texttt{MAPP2}~\cite{Pinfold:2019nqj,Pinfold:2019zwp}, and \texttt{MATHUSLA}~\cite{Chou:2016lxi,Curtin:2018mvb,MATHUSLA:2018bqv,MATHUSLA:2020uve,MATHUSLA:2025zyt}.
The \texttt{SHiP} experiment has been approved in 2024 and is currently planned to begin {commissioning and first data-taking between 2031 and 2033~\cite{cern_medium_term_plan,SHiP:2025ows}}.
Sensitivity projections for these experiments across a wide range of LLP scenarios have been extensively studied; see e.g., Refs.~\cite{Beltran:2025oqj,Patrone:2025fwk,Ema:2025bww,Wang:2024mrc,A:2025ygb,Ahmed:2025ldh} for recent phenomenological works.

Very recently, the overall plans of several of these experiments have undergone major revisions, most notably for \texttt{MATHUSLA}~\cite{MATHUSLA:2025zyt} and \texttt{ANUBIS}~\cite{ANUBIS:2025sgg}.
The planned fiducial volume of \texttt{MATHUSLA} has been substantially reduced, and the proposed location of the \texttt{ANUBIS} detector has been changed from a service shaft to the cavern hosting the \texttt{ATLAS} detector, with the \texttt{ANUBIS} detector installed on the cavern ceiling.
The \texttt{SHiP} experiment employs its Hidden Sector Decay Spectrometer (\texttt{HSDS}) as the decay volume for LLPs, and while its geometry has also been updated, the changes are comparatively modest.
In addition, its projected operational duration has been updated to 15 years, which we adopt in this work.

For most of the experiments considered here, the large distances from the IPs allow for the installation of shielding and veto systems, leading to expectations of negligible background levels.
An important exception is \texttt{ANUBIS}, for which the revised detector design implies significantly different background conditions.
These have been carefully evaluated in Ref.~\cite{ANUBIS:2025sgg}.\footnote{A complementary discussion of background estimates at \texttt{ANUBIS} can be found in Appendix~B of Ref.~\cite{Beltran:2025oqj}.}
The updated background levels have a substantial impact on the sensitivity reach and are therefore taken into account in our numerical analysis.

Taken together, these recent developments motivate a timely re-assessment of the sensitivity projections for LLP searches at the \texttt{LHC} far detectors and \texttt{SHiP}, as well as a re-evaluation of their relative performance.

A wide variety of BSM scenarios predict LLPs.
In this work, we focus on heavy neutral leptons (HNLs)~\cite{Shrock:1980vy,Shrock:1980ct,Shrock:1981wq}, also commonly referred to right-handed neutrinos, heavy neutrinos, or sterile neutrinos.\footnote{See Ref.~\cite{Abdullahi:2022jlv} for a recent overview of the present and future status of the HNLs.}
HNLs are SM-singlet fermions that mix with active neutrinos.
While HNLs can also appear in non-minimal scenarios involving additional BSM degrees of freedom~\cite{Mohapatra:1974gc,Bando:1998ww,Chiang:2019ajm,Dorsner:2016wpm}, here we focus on the minimal scenario in which HNLs constitute the only new states. This provides a clean benchmark for comparing detector sensitivities without introducing additional model-dependent parameters.
Although many models predict three generations of HNLs, we assume for simplicity that only a single HNL, denoted by $N$, is kinematically accessible, with the remaining states sufficiently heavy to decouple.
We further assume that $N$ mixes exclusively with the electron neutrino.
This benchmark scenario has been widely adopted in studies of the experimental sensitivity to HNLs.
Finally, we restrict our attention to HNLs produced in decays of heavy mesons.
For existing studies on the sensitivity reach of the \texttt{LHC} far detectors and \texttt{SHiP} to the minimal HNLs, see, e.g., Refs.~\cite{Curtin:2018mvb,SHiP:2018xqw,FASER:2018eoc, Kling:2018wct,Helo:2018qej,CODEX-b:2019jve,FASER:2019aik,Hirsch:2020klk,DeVries:2020jbs,Ovchynnikov:2022its,Gunther:2023vmz}; see also early proposals to search for HNLs at accelerator-based experiments in e.g.~Refs.~\cite{Fargion:1995qb,Fargion:1999ss}.

\begin{figure}[t]
    \centering
    \includegraphics[width=\linewidth]{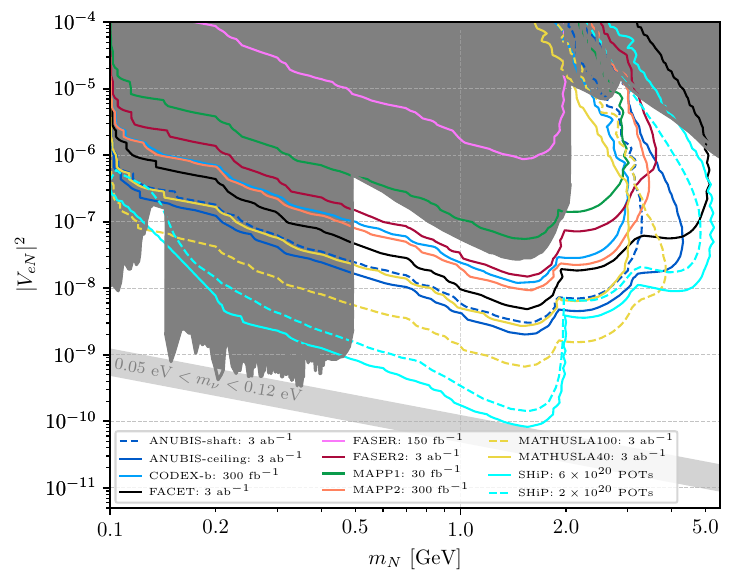}
    \caption{Updated sensitivity reach of the \texttt{LHC} far detectors and the \texttt{SHiP} beam-dump experiment to minimal heavy neutral leptons that mix with the electron neutrino only, shown in the $(m_N, |V_{eN}|^2)$ plane. The dark gray area represents the current experimental bounds obtained at \texttt{PIENU}~\cite{PIENU:2017wbj}, \texttt{KENU}~\cite{Bryman:2019bjg}, \texttt{CHARM}~\cite{CHARM:1985nku}, \texttt{NA62}~\cite{NA62:2020mcv,NA62:2025csa}, \texttt{T2K}~\cite{T2K:2019jwa}, \texttt{BEBC}~\cite{Barouki:2022bkt},  \texttt{DELPHI}~\cite{DELPHI:1996qcc}, \texttt{ATLAS}~\cite{ATLAS:2022atq}, and \texttt{CMS}~\cite{CMS:2024ake}. The light gray band corresponds to the parameter region targeted by the type-I seesaw mechanism for active-neutrino masses between 0.05~eV and 0.12~eV. The sensitivity in the low-mass region ($m_N \lesssim m_D$) is dominated by HNLs produced from charm-meson decays, whereas for heavier masses ($m_D \lesssim m_N \lesssim m_B$) the contribution from bottom-meson decays dominates; only $D$- and $B$-meson channels are included in this analysis. The sensitivities of the outdated designs are displayed as dashed lines.}
    \label{fig:sensitivity}
\end{figure}

In this work, we implement the latest designs of the experiments discussed above in the public tool Displaced Decay Counter (\texttt{DDC})~\cite{Domingo:2023dew,DDC_github} to compute the acceptances of the \texttt{LHC} far detectors and \texttt{SHiP} for long-lived HNLs.
Based on these acceptances, we evaluate the corresponding sensitivity reach, shown in Fig.~\ref{fig:sensitivity}.
{These updated results show, as expected, important changes in the sensitivity projections for \texttt{MATHUSLA} and \texttt{ANUBIS} particularly, comparing their latest designs with their previous configurations.}
Details of the experimental configurations and the simulation procedure underlying these results are described in the following sections.
These results provide a useful reference for comparing the constraining power of the considered experiments for LLPs produced in heavy-meson decays at the \texttt{LHC} and at \texttt{SHiP}.

\vspace{0.2cm}
\noindent \textbf{Model.} After spontaneous electroweak symmetry breaking, the relevant interactions of the heavy neutral lepton are described by the Lagrangian
\begin{eqnarray}
		\mathcal{L}_{N} &=& \frac{g}{\sqrt{2}}\ \sum_{\alpha}
		V_{\alpha N}\,\bar \ell_\alpha \gamma^{\mu} P_L N\,W^-_{L \mu}\nonumber\\
		&&+\frac{g}{2 \cos\theta_W}\ \sum_{\alpha, i}V^{L}_{\alpha i} V_{\alpha N}^* \,\overline{N} \gamma^{\mu} P_L \nu_{i}\, Z_{\mu},
		\label{eqn:Lagrangian}
\end{eqnarray}
where $g$ denotes the SU(2) gauge coupling, $V_{\alpha N}$ parameterizes the mixing between the active neutrino $\nu_\alpha$ and the HNL $N$ with $\alpha=e, \mu, \tau$, $V^L$ is the left-handed neutrino mixing matrix, and $\theta_W$ is the Weinberg angle.

We assume the HNL to be a Majorana fermion.
HNLs can be produced in rare, prompt, leptonic and semi-leptonic decays of charm and bottom mesons,\footnote{We neglect HNL production from pion, kaons, and heavier mesons in this work.} and subsequently decay via both charged-current and neutral-current interactions into lighter SM particles.
Throughout this work, we follow Ref.~\cite{DeVries:2020jbs} for the calculation of meson decay branching ratios (BRs) into HNLs, as well as for the computation of HNL decay widths.
We will study the range of HNL masses between 0.1~GeV and the $B$-meson thresholds.

\vspace{0.2cm}
\noindent \textbf{Experiments \& simulation.} As discussed in the first section, we study \texttt{LHC} far-detector experiments and the \texttt{SHiP} beam-dump experiment.
We will first elaborate on the experiments of which the geometrical designs have been significantly changed and hence for which the discovery prospects of LLPs should be largely modified.
These experiments are \texttt{MATHUSLA} and \texttt{ANUBIS}.

\texttt{MATHUSLA} was initially proposed as a rectangular box with the dimensions of 200~m$\times$200~m$\times$20~m for its fiducial volume, to be 100~m distanced both vertically ($y$-axis) and horizontally ($z$-axis) from the ATLAS or CMS IP~\cite{Chou:2016lxi,Curtin:2018mvb,MATHUSLA:2018bqv}.
In its updated Letter of Intent~\cite{MATHUSLA:2020uve}, the geometrical design has changed to 100~m$\times$100~m$\times$25~m, positions with a horizontal (vertical) distance of 68~m (60~m) from the CMS IP; we coin this configuration as \texttt{MATHUSLA100}.
In the lately published Conceptual Design Report for \texttt{MATHUSLA}~\cite{MATHUSLA:2025zyt}, for cost reasons the fiducial volume has further shrunk, now to 40~m$\times$40~m$\times$11~m (\texttt{MATHUSLA40}).
It is supposed to be 70~m and 81~m away from the CMS IP in the $z$- and $y$-directions, respectively.

The geometrical configuration of \texttt{ANUBIS} has been critically updated, too.
Originally proposed to be installed inside one of the service shafts above the \texttt{ATLAS} IP, \texttt{ANUBIS-shaft}~\cite{Bauer:2019vqk}\footnote{See also Ref.~\cite{Hirsch:2020klk} for one of the first LLP sensitivity studies of  \texttt{ANUBIS-shaft}.} should be 5~m and 24~m away from the \texttt{ATLAS} IP in $z$- and $y$-directions.
It is cylindrically shaped with a height of 56~m and a diameter of 18~m.
The LLPs are supposed to decay inside this cylinder with the charged decay products to hit on three track stations inside.
The latest design of \texttt{ANUBIS}~\cite{ANUBIS:2025sgg} takes a distinct direction, now exploiting the space between the \texttt{ATLAS} main detector and the cavern ceiling as the fiducial volume.
It has a shape of an annular cylindrical sector of length 53~m; its inner and outer radii are 11.3~m and 19.3~m; and it has an azimuthal coverage of $\sim 20\%$.
Compared to \texttt{ANUBIS-shaft}, \texttt{ANUBIS-ceiling} is closer to the IP, guaranteeing both better acceptance rates and exacerbated background levels.

Besides \texttt{MATHUSLA} and \texttt{ANUBIS}, \texttt{SHiP} has its designs updated, though not to a large degree.
We only list the geometrical dimensions of the latest design here, extracted from Ref.~\cite{Albanese:2878604}.\footnote{We refer to Refs.~\cite{SHiP:2018yqc,DeVries:2020jbs} for detail of the previous geometrical design of \texttt{SHiP}, which is not largely different from the latest design.}
The near end of the \texttt{SHiP-HSDS} which is the decay volume for the LLPs at the \texttt{SHiP} experiment, is 33~m away behind the target.
The \texttt{HSDS} has a length of 50~m, and its front (rear) face has dimensions $(x,y)=1.0~\text{m}\times 2.7~\text{m}$ $(4.0~\text{m}\times 6.0~\text{m})$.
It is worth mentioning that compared to the relatively mild updates in its geometries, the \texttt{SHiP} experiment has its primary update in its projected operation period, now prolonged from 5 years to 15 years.

We also refer to the tool \texttt{SensCalc}~\cite{Ovchynnikov:2023cry} which, similarly, includes the implementations of \texttt{ANUBIS-shaft}, \texttt{ANUBIS-ceiling}, and the latest configuration of \texttt{SHiP}.
The geometrical setups of these experiments in \texttt{DDC}, as we have checked, are in agreement with those in \texttt{SensCalc}.

For summaries of the other proposals considered in this work, we refer to e.g., Refs.~\cite{Wang:2024mrc,Gunther:2023vmz,DeVries:2020jbs}.
They have all been implemented in \texttt{DDC} and in this work we rely on \texttt{DDC} to derive their sensitivity reach.

As mentioned above, we will focus on HNLs produced from rare decays of charm and bottom mesons.
Concretely, we will consider the $D^0, D^+, D_s^+, B^0, B^+, B_s$ mesons, as well as their charge-conjugated counterparts.
The production rates of these mesons, estimated for a $4\pi$ solid-angle coverage~\cite{DeVries:2020jbs,Bondarenko:2018ptm}, are reproduced here:
\begin{eqnarray}
    N_{D^0+\overline{D}^0}^\text{\texttt{HL-LHC}}=3.89\times 10^{16}, N_{D^++D^-}^\text{\texttt{HL-LHC}}=2.04\times 10^{16}, \nonumber\\
    N_{D_s^++D_s^-}^\text{\texttt{HL-LHC}}=6.62\times 10^{15}, N_{D^0+\overline{D}^0}^\text{\texttt{SHiP}}=1.29\times 10^{18},\nonumber\\
    N_{D^++D^-}^\text{\texttt{SHiP}}=4.2\times 10^{17}, N_{D_s^++D_s^-}^\text{\texttt{SHiP}}=1.8\times 10^{17},\nonumber\\
    N_{B^0+\overline{B}^0}^\text{\texttt{HL-LHC}}=1.46\times 10^{15}, N_{B^++B^-}^\text{\texttt{HL-LHC}}=1.46\times 10^{15},\nonumber\\
    N_{B_s0++\overline{B}_s^0}^\text{\texttt{HL-LHC}}=2.53\times 10^{14}, N_{B^0+\overline{B}^0}^\text{\texttt{SHiP}}=8.1\times 10^{13},\nonumber\\
    N_{B^++B^-}^\text{\texttt{SHiP}}=8.1\times 10^{13}, N_{B_s^0+\overline{B}_s^0}^\text{\texttt{SHiP}}=2.16\times 10^{13}, \label{eqn:NM}
\end{eqnarray}
where we assume an integrated luminosity of 3~ab$^{-1}$ for the \texttt{HL-LHC} and a nominal 15-year operation of \texttt{SHiP} corresponding to $6\times 10^{20}$ protons on target (POTs).\footnote{In the previous design of the \texttt{SHiP} experiment~\cite{SHiP:2018yqc}, a period of 5 years corresponding to $2\times 10^{20}$ POTs was planned.}

We then implement the latest geometries of \texttt{MATHUSLA40}, \texttt{ANUBIS-ceiling}, as well as \texttt{SHiP-HSDS} into the tool \texttt{DDC} which is then ready for calculating the acceptances of all the \texttt{LHC} far detectors and \texttt{SHiP} to the minimal HNLs.
We proceed by computing the signal-event rates with the following expressions,
\begin{eqnarray}
    N_S^{N,D}&=&\sum_{M_D} N_{M_D} \cdot \mathcal{B}(M_D\to N + X) \cdot \epsilon_D \cdot\mathcal{B}(N\to \text{vis.}),\quad\\
    N_S^{N,B}&=&\sum_{M_B} N_{M_B} \cdot \mathcal{B}(M_B\to N + X) \cdot \epsilon_B \cdot \mathcal{B}(N\to \text{vis.}),\quad
\end{eqnarray}
where we have split the computation into two parts, for $D$-meson and $B$-meson decays, respectively, and assumed that the kinematics of the HNLs produced from different charm mesons are approximately identical (the same for the different bottom mesons).
The number of $D$- and $B$-mesons $N_{M_D}$ and $N_{M_B}$ are given in Eq.~\eqref{eqn:NM}, and $\mathcal{B}(N\to\text{vis.})$ is the decay branching ratio of the HNL into visible final states for which we assume all the decay channels of $N$ except the fully invisible tri-neutrino one are detectable.
Finally, $\epsilon_D$ and $\epsilon_B$ are estimated with \texttt{DDC}.

\texttt{DDC} loads \texttt{Pythia8}~\cite{Sjostrand:2014zea} for simulating hard QCD $c\bar{c}$ and $b\bar{b}$ events.
\texttt{Pythia8} performs showering and hadronization, and the simulated $D$- and $B$-mesons are set to decay exclusively into the HNL with the relative branching ratios between the different channels aligned with the meson-decay BR computation results.
\texttt{DDC} then takes into account the kinematics of the HNLs provided by \texttt{Pythia8}, the geometries and positions of the considered detectors, as well as the HNL masses, and makes use of exponential decay distribution formula to calculate the acceptances:
\begin{eqnarray}
\epsilon_{D/B} &=& \frac{1}{N_{\text{MC}}}\sum_{i=1}^{N_{\text{MC}}}\epsilon_i,\\
\epsilon_i &=& \frac{\delta\phi}{2\pi}\, e^{\Big(  -\frac{D}{\beta_i\gamma_i c \tau_N} \Big)}\,\Big(1-e^{-\frac{L}{\beta_i\gamma_i c \tau_N}}\Big),
\end{eqnarray}
where $N_{\text{MC}}$ is the total number of Monte-Carlo simulation events (we have simulated ten thousand or one hundred thousand events for each mass point in different cases), $\delta\phi$ is the azimuthal-angle coverage of the detector, $D$ is the distance between the hard-collision point and the detector along the traveling direction of the HNL, $L$ is the distance that the HNL would travel inside the detector if it does not decay before leaving the detector, $\beta_i$ and $\gamma_i$ are the speed and Lorentz boost factor of the $i^{\text{th}}$ simulated HNL, $c$ labels the speed of light, and $\tau_N$ is the HNL's mean lifetime.
We assume 100\% detection efficiency for the visible decays.
For more detail of the computation procedure applied in \texttt{DDC} for the different detectors, see Ref.~\cite{Domingo:2023dew}.

As discussed earlier, we would assume zero background for all the experiments except \texttt{ANUBIS-shaft} and \texttt{ANUBIS-ceiling}.
The background levels of LLP searches at \texttt{ANUBIS-ceiling} and \texttt{ANUBIS-shaft} have been estimated in Ref.~\cite{ANUBIS:2025sgg}, which reports $182.4\pm 12.2$ and $63.7\pm4.3$, respectively.
Following Ref.~\cite{Hirsch:2020klk} we take $N_{S95}^{\texttt{ANUBIS-ceiling}}=2\times \sqrt{195}\approx 28$ and $N_{S95}^{\texttt{ANUBIS-shaft}}=2\times \sqrt{68}\approx 17$ as the signal-event numbers corresponding to the exclusion limits at 95\% confidence level (C.L.) for the two \texttt{ANUBIS} configurations, respectively.
For the zero-background experiments, we show contour curves for 3 signal events as the sensitivity reach at 95\% C.L.

\vspace{0.2cm}
\noindent \textbf{Results.} The numerical results are displayed in Fig.~\ref{fig:sensitivity} in the $(m_N,|V_{eN}|^2)$ plane where the contour curves are exclusion boundaries at 95\% C.L.
For each experiment, the sensitivities are dominated by the charm-meson (bottom-meson) contributions for $m_N\lesssim m_D$ ($m_{D}\lesssim m_N \lesssim m_B$), mainly as a result of the charm-meson productions rates several orders of magnitude larger than those of the bottom mesons.
The sensitivity reaches of the up-to-date designs are shown with solid lines, while those of the outdated ones with dashed curves.
We find that for $m_N\lesssim m_D$ the sensitivity reach is dominated by \texttt{SHiP}, followed by \texttt{MATHUSLA100} and \texttt{ANUBIS-ceiling}, while for heavier HNLs \texttt{MATHUSLA100} and again \texttt{SHiP} are expected to dominate the sensitivity reach.
We note that the sensitivity results of the non-updated LLP experiments (such as \texttt{FASER} and \texttt{CODEX-b}) that we obtained with \texttt{DDC} are, as we checked, in excellent agreement with those reported in the existing literature~\cite{Curtin:2018mvb,SHiP:2018xqw,FASER:2018eoc, Kling:2018wct,Helo:2018qej,CODEX-b:2019jve,FASER:2019aik,Hirsch:2020klk,DeVries:2020jbs,Ovchynnikov:2022its,Gunther:2023vmz}.

The dark-gray regions have been excluded, representing a combination of the constraints obtained at \texttt{PIENU}~\cite{PIENU:2017wbj}, \texttt{KENU}~\cite{Bryman:2019bjg}, \texttt{CHARM}~\cite{CHARM:1985nku}, \texttt{NA62}~\cite{NA62:2020mcv,NA62:2025csa}, \texttt{T2K}~\cite{T2K:2019jwa}, \texttt{BEBC}~\cite{Barouki:2022bkt}, \texttt{DELPHI}~\cite{DELPHI:1996qcc}, \texttt{ATLAS}~\cite{ATLAS:2022atq}, and \texttt{CMS}~\cite{CMS:2024ake}.
Also, we present a band in light gray, corresponding to the parameter region targeted by the type-I seesaw relation, $|V_{eN}|^2\simeq m_{\nu}/m_N$, for active-neutrino masses between 0.05~eV and 0.12~eV; the two values of the active-neutrino masses are inspired by neutrino-oscillations observations~\cite{Canetti:2010aw} and cosmological observations~\cite{Planck:2018vyg}, respectively.
In general we find almost all of the considered experiments can probe large parameter regions beyond the present bounds.
\texttt{FASER} cannot test any unexcluded parameter region mainly because of its relatively small volume and limited dataset it is expected to receive ($150~\text{fb}^{-1}$).

As expected from considerations of detector volumes, \texttt{MATHUSLA40} is found to only probe $|V_{eN}|^2$ values larger than \texttt{MATHUSLA100} can by a factor of about 5.
Meanwhile, its exclusion limits turn out to agree well with those of \texttt{ANUBIS-shaft} except when $m_N\gtrsim 2.8~\text{GeV}$ where \texttt{ANUBIS-shaft} performs better mainly because it is more transverse.
\texttt{ANUBIS-ceiling} is closer to the ATLAS IP than \texttt{ANUBIS-shaft}, thus resulting in both higher background levels and enhanced acceptances.
Our numerical simulations show that \texttt{ANUBIS-ceiling} can test the mixing parameter squared smaller than \texttt{ANUBIS-shaft} can by approximately 2.
Also, the latest design of the \texttt{SHiP} experiment is found to outperform the previous one by roughly a factor of 2 in testing $|V_{eN}|^2$, which is dominantly the consequence of the enhanced POT rates.

We note that while we have confined ourselves to considering an HNL mixed with the electron neutrino only, sensitivities for an HNL mixed with the muon neutrino only would be quite similar except for some minor effects of kinematic thresholds.
As for an HNL mixed solely with the tau neutrino, tangibly reduced sensitivity reach is expected, because of relatively stronger suppression effects of kinematic thresholds and worse reconstruction efficiencies of the $\tau$-leptons.

In addition, in the present paper we have assumed a Majorana HNL; if the HNL is of Dirac nature, the decay widths of the HNL should be reduced by a factor of 2 and the lower sensitivity reach of the studied experiments is correspondingly weakened by $\sim\sqrt{2}$.

\vspace{0.2cm}
\noindent \textbf{Summary.} In this work, we have implemented the updated geometrical designs of \texttt{MATHUSLA}, \texttt{ANUBIS}, and \texttt{SHiP} in the public tool \texttt{DDC} to compute the signal acceptances and event yields for a long-lived heavy neutral lepton mixed with the electron neutrino only.
In particular, we incorporate the recently updated background estimates for \texttt{ANUBIS} reported in Ref.~\cite{ANUBIS:2025sgg}, going beyond the commonly adopted optimistic assumption of vanishing background, and account for the latest projected operational duration of the \texttt{SHiP} beam-dump experiment of 15 years.

Compared to \texttt{MATHUSLA100}, the reduced fiducial volume of \texttt{MATHUSLA40} leads to a sensitivity to the mixing parameter squared $|V_{eN}|^2$ weakened by a factor of $\sim 5$, rendering it broadly competitive with the \texttt{ANUBIS-shaft} configuration except in the high-mass regime.
The \texttt{ANUBIS-ceiling} setup, benefiting from its closer proximity to the IP, is expected to probe values of $|V_{eN}|^2$ smaller than those accessible to \texttt{ANUBIS-shaft} by a factor of $\sim 2$.
The latest plan of \texttt{SHiP} can exclude $|V_{eN}|^2$ values smaller than those the previous plan can by a factor of about 2, primarily owing to the relatively modest changes in its geometries and the currently 3-times prolonged operation period.
The \texttt{SHiP} experiment, with its latest design, is found to dominate the sensitivity reach for $m_N$ just below the charm threshold and in the mass window $3.8~\text{GeV}\lesssim m_N\lesssim 5.2~\text{GeV}$, while \texttt{MATHUSLA100} provides the strongest reach in the intermediate mass range.

With this short paper, we provide the LLP community with a concise, up-to-date reference on the sensitivity reach of the \texttt{LHC} far detectors and \texttt{SHiP} to one of the most widely studied LLP scenarios, namely minimal heavy neutral leptons.
While our study focuses on minimal HNLs, the same procedure can readily be applied to non-minimal HNL scenarios or other LLP models, allowing rapid assessment of the updated detector designs studied here for a broader class of BSM signatures.

\vspace{0.5cm}
\noindent \textbf{Acknowledgments.} Z.S.W.~would like to thank Martin Hirsch for introducing him to the study of long-lived particles, including heavy neutral leptons, and for eight years of close collaboration.
This work was supported by the National Natural Science Foundation of China under grant Nos.~12475106 and 12505120, and the Fundamental Research Funds for the Central Universities under Grant No.~JZ2025HGTG0252.



\bibliography{refs}

@article{ANUBIS:2025sgg,
    author = "Brandt, Oleg and others",
    collaboration = "ANUBIS",
    title = "{The ANUBIS detector and its sensitivity to neutral long-lived particles}",
    eprint = "2510.26932",
    archivePrefix = "arXiv",
    primaryClass = "hep-ex",
    month = "10",
    year = "2025"
}

@article{A:2025ygb,
    author = "A, ShivaSankar K. and Das, Souvik and Das, Arindam and Mandal, Sanjoy",
    title = "{Right handed neutrino production from $Z^\prime$ interactions in forward search experiments}",
    eprint = "2508.10734",
    archivePrefix = "arXiv",
    primaryClass = "hep-ph",
    month = "8",
    year = "2025"
}

@inproceedings{Alimena:2025kjv,
    author = "Alimena, J. and others",
    title = "{Feebly-Interacting Particles: FIPs at LHCb {\textemdash} Workshop Report 2025 Edition}",
    booktitle = "{LHCb FIP Physics Workshop 2025}",
    eprint = "2510.05257",
    archivePrefix = "arXiv",
    primaryClass = "hep-ph",
    month = "10",
    year = "2025"
}

@article{Ahmed:2025ldh,
    author = "Ahmed, Aqeel and Chacko, Zackaria and Desai, Niral and Doshi, Sanket and Kilic, Can and Najjari, Saereh and Sudha, Ram Purandhar Reddy",
    title = "{Long-Lived-Particle Signals of a Composite Hidden Sector through the Neutrino Portal}",
    eprint = "2512.09046",
    archivePrefix = "arXiv",
    primaryClass = "hep-ph",
    reportNumber = "UT-WI-28-2025",
    month = "12",
    year = "2025"
}

@article{Canetti:2010aw,
    author = "Canetti, Laurent and Shaposhnikov, Mikhail",
    title = "{Baryon Asymmetry of the Universe in the NuMSM}",
    eprint = "1006.0133",
    archivePrefix = "arXiv",
    primaryClass = "hep-ph",
    doi = "10.1088/1475-7516/2010/09/001",
    journal = "JCAP",
    volume = "09",
    pages = "001",
    year = "2010"
}

@article{Planck:2018vyg,
    author = "Aghanim, N. and others",
    collaboration = "Planck",
    title = "{Planck 2018 results. VI. Cosmological parameters}",
    eprint = "1807.06209",
    archivePrefix = "arXiv",
    primaryClass = "astro-ph.CO",
    doi = "10.1051/0004-6361/201833910",
    journal = "Astron. Astrophys.",
    volume = "641",
    pages = "A6",
    year = "2020",
    note = "[Erratum: Astron.Astrophys. 652, C4 (2021)]"
}

@article{Domingo:2023dew,
    author = {Domingo, Florian and G{\"u}nther, Julian and Kim, Jong Soo and Wang, Zeren Simon},
    title = "{A C++ program for estimating detector sensitivities to long-lived particles: displaced decay counter}",
    eprint = "2308.07371",
    archivePrefix = "arXiv",
    primaryClass = "hep-ph",
    doi = "10.1140/epjc/s10052-024-13009-9",
    journal = "Eur. Phys. J. C",
    volume = "84",
    number = "6",
    pages = "642",
    year = "2024"
}

@article{PIENU:2017wbj,
    author = "Aguilar-Arevalo, A. and others",
    collaboration = "PIENU",
    title = "{Improved search for heavy neutrinos in the decay $\pi\rightarrow e\nu$}",
    eprint = "1712.03275",
    archivePrefix = "arXiv",
    primaryClass = "hep-ex",
    doi = "10.1103/PhysRevD.97.072012",
    journal = "Phys. Rev. D",
    volume = "97",
    number = "7",
    pages = "072012",
    year = "2018"
}

@article{NA62:2020mcv,
    author = "Cortina Gil, Eduardo and others",
    collaboration = "NA62",
    title = "{Search for heavy neutral lepton production in $K^+$ decays to positrons}",
    eprint = "2005.09575",
    archivePrefix = "arXiv",
    primaryClass = "hep-ex",
    reportNumber = "CERN-EP-2020-089",
    doi = "10.1016/j.physletb.2020.135599",
    journal = "Phys. Lett. B",
    volume = "807",
    pages = "135599",
    year = "2020"
}

@article{T2K:2019jwa,
    author = "Abe, K. and others",
    collaboration = "T2K",
    title = "{Search for heavy neutrinos with the T2K near detector ND280}",
    eprint = "1902.07598",
    archivePrefix = "arXiv",
    primaryClass = "hep-ex",
    doi = "10.1103/PhysRevD.100.052006",
    journal = "Phys. Rev. D",
    volume = "100",
    number = "5",
    pages = "052006",
    year = "2019"
}

@article{Barouki:2022bkt,
    author = "Barouki, Ryan and Marocco, Giacomo and Sarkar, Subir",
    title = "{Blast from the past II: Constraints on heavy neutral leptons from the BEBC WA66 beam dump experiment}",
    eprint = "2208.00416",
    archivePrefix = "arXiv",
    primaryClass = "hep-ph",
    doi = "10.21468/SciPostPhys.13.5.118",
    journal = "SciPost Phys.",
    volume = "13",
    pages = "118",
    year = "2022"
}

@article{DELPHI:1996qcc,
    author = "Abreu, P. and others",
    collaboration = "DELPHI",
    title = "{Search for neutral heavy leptons produced in Z decays}",
    reportNumber = "CERN-PPE-96-195",
    doi = "10.1007/s002880050370",
    journal = "Z. Phys. C",
    volume = "74",
    pages = "57--71",
    year = "1997",
    note = "[Erratum: Z.Phys.C 75, 580 (1997)]"
}

@article{CHARM:1985nku,
    author = "Bergsma, F. and others",
    collaboration = "CHARM",
    title = "{A Search for Decays of Heavy Neutrinos in the Mass Range 0.5-{GeV} to 2.8-{GeV}}",
    reportNumber = "CERN-EP-85-190",
    doi = "10.1016/0370-2693(86)91601-1",
    journal = "Phys. Lett. B",
    volume = "166",
    pages = "473--478",
    year = "1986"
}

@article{Bryman:2019bjg,
    author = "Bryman, D. A. and Shrock, R.",
    title = "{Constraints on Sterile Neutrinos in the MeV to GeV Mass Range}",
    eprint = "1909.11198",
    archivePrefix = "arXiv",
    primaryClass = "hep-ph",
    reportNumber = "TRIUMF-UBC-Stony Brook preprint (YITP-SB-2019-9)",
    doi = "10.1103/PhysRevD.100.073011",
    journal = "Phys. Rev. D",
    volume = "100",
    pages = "073011",
    year = "2019"
}

@article{Bauer:2019vqk,
    author = "Bauer, Martin and Brandt, Oleg and Lee, Lawrence and Ohm, Christian",
    title = "{ANUBIS: Proposal to search for long-lived neutral particles in CERN service shafts}",
    eprint = "1909.13022",
    archivePrefix = "arXiv",
    primaryClass = "physics.ins-det",
    month = "9",
    year = "2019"
}

@article{Feng:2017uoz,
    author = "Feng, Jonathan L. and Galon, Iftah and Kling, Felix and Trojanowski, Sebastian",
    title = "{ForwArd Search ExpeRiment at the LHC}",
    eprint = "1708.09389",
    archivePrefix = "arXiv",
    primaryClass = "hep-ph",
    reportNumber = "UCI-TR-2017-08",
    doi = "10.1103/PhysRevD.97.035001",
    journal = "Phys. Rev. D",
    volume = "97",
    number = "3",
    pages = "035001",
    year = "2018"
}

@article{FASER:2018eoc,
    author = "Ariga, Akitaka and others",
    collaboration = "FASER",
    title = "{FASER\textquoteright{}s physics reach for long-lived particles}",
    eprint = "1811.12522",
    archivePrefix = "arXiv",
    primaryClass = "hep-ph",
    reportNumber = "UCI-TR-2018-19, KYUSHU-RCAPP-2018-06",
    doi = "10.1103/PhysRevD.99.095011",
    journal = "Phys. Rev. D",
    volume = "99",
    number = "9",
    pages = "095011",
    year = "2019"
}

@article{Cerci:2021nlb,
    author = "Cerci, S. and others",
    title = "{FACET: A new long-lived particle detector in the very forward region of the CMS experiment}",
    eprint = "2201.00019",
    archivePrefix = "arXiv",
    primaryClass = "hep-ex",
    doi = "10.1007/JHEP06(2022)110",
    journal = "JHEP",
    volume = "06",
    pages = "110",
    year = "2022"
}

@article{Pinfold:2019nqj,
    author = "Pinfold, James Lewis",
    doi = "10.3390/universe5020047",
    journal = "Universe",
    number = "2",
    pages = "47",
    title = "{The MoEDAL Experiment at the LHC---A Progress Report}",
    volume = "5",
    year = "2019"
}

@article{Pinfold:2019zwp,
    author = "Pinfold, James L.",
    editor = "Dainton, John",
    doi = "10.1098/rsta.2019.0382",
    journal = "Phil. Trans. Roy. Soc. Lond. A",
    number = "2161",
    pages = "20190382",
    title = "{The MoEDAL experiment: a new light on the high-energy frontier}",
    volume = "377",
    year = "2019"
}

@article{Gligorov:2017nwh,
    author = "Gligorov, Vladimir V. and Knapen, Simon and Papucci, Michele and Robinson, Dean J.",
    title = "{Searching for Long-lived Particles: A Compact Detector for Exotics at LHCb}",
    eprint = "1708.09395",
    archivePrefix = "arXiv",
    primaryClass = "hep-ph",
    doi = "10.1103/PhysRevD.97.015023",
    journal = "Phys. Rev. D",
    volume = "97",
    number = "1",
    pages = "015023",
    year = "2018"
}

@article{Aielli:2019ivi,
    author = "Aielli, Giulio and others",
    title = "{Expression of interest for the CODEX-b detector}",
    eprint = "1911.00481",
    archivePrefix = "arXiv",
    primaryClass = "hep-ex",
    doi = "10.1140/epjc/s10052-020-08711-3",
    journal = "Eur. Phys. J. C",
    volume = "80",
    number = "12",
    pages = "1177",
    year = "2020"
}

@article{SHiP:2015vad,
    author = "Anelli, M. and others",
    collaboration = "SHiP",
    title = "{A facility to Search for Hidden Particles (SHiP) at the CERN SPS}",
    eprint = "1504.04956",
    archivePrefix = "arXiv",
    primaryClass = "physics.ins-det",
    reportNumber = "CERN-SPSC-2015-016, SPSC-P-350",
    month = "4",
    year = "2015"
}

@article{Alekhin:2015byh,
    author = "Alekhin, Sergey and others",
    title = "{A facility to Search for Hidden Particles at the CERN SPS: the SHiP physics case}",
    eprint = "1504.04855",
    archivePrefix = "arXiv",
    primaryClass = "hep-ph",
    reportNumber = "CERN-SPSC-2015-017, SPSC-P-350-ADD-1",
    doi = "10.1088/0034-4885/79/12/124201",
    journal = "Rept. Prog. Phys.",
    volume = "79",
    number = "12",
    pages = "124201",
    year = "2016"
}

@article{SHiP:2018xqw,
    author = "Ahdida, C. and others",
    collaboration = "SHiP",
    title = "{Sensitivity of the SHiP experiment to Heavy Neutral Leptons}",
    eprint = "1811.00930",
    archivePrefix = "arXiv",
    primaryClass = "hep-ph",
    doi = "10.1007/JHEP04(2019)077",
    journal = "JHEP",
    volume = "04",
    pages = "077",
    year = "2019"
}

@article{SHiP:2021nfo,
    author = "Ahdida, C. and others",
    collaboration = "SHiP",
    title = "{The SHiP experiment at the proposed CERN SPS Beam Dump Facility}",
    eprint = "2112.01487",
    archivePrefix = "arXiv",
    primaryClass = "physics.ins-det",
    doi = "10.1140/epjc/s10052-022-10346-5",
    journal = "Eur. Phys. J. C",
    volume = "82",
    number = "5",
    pages = "486",
    year = "2022"
}

@article{MATHUSLA:2025zyt,
    author = "Aitken, Branden and others",
    collaboration = "MATHUSLA",
    title = "{Conceptual Design Report for the MATHUSLA Long-Lived Particle Detector near CMS}",
    eprint = "2503.20893",
    archivePrefix = "arXiv",
    primaryClass = "physics.ins-det",
    reportNumber = "FERMILAB-PUB-25-0314-PPD",
    month = "3",
    year = "2025"
}

@techreport{Albanese:2878604,
      author        = "Albanese, R and others",
      collaboration = "SHiP",
      title         = "{BDF/SHiP at the ECN3 high-intensity beam facility}",
      institution   = "CERN",
      reportNumber  = "CERN-SPSC-2023-033, SPSC-P-369",
      address       = "Geneva",
      year          = "2023, CERN-SPSC-2023-033, SPSC-P-369",
      url           = "https://cds.cern.ch/record/2878604",
}

@article{Curtin:2018mvb,
    author = "Curtin, David and others",
    title = "{Long-Lived Particles at the Energy Frontier: The MATHUSLA Physics Case}",
    eprint = "1806.07396",
    archivePrefix = "arXiv",
    primaryClass = "hep-ph",
    reportNumber = "FERMILAB-PUB-18-264-T",
    doi = "10.1088/1361-6633/ab28d6",
    journal = "Rept. Prog. Phys.",
    volume = "82",
    number = "11",
    pages = "116201",
    year = "2019"
}

@article{MATHUSLA:2020uve,
    author = "Alpigiani, Cristiano and others",
    collaboration = "MATHUSLA",
    title = "{An Update to the Letter of Intent for MATHUSLA: Search for Long-Lived Particles at the HL-LHC}",
    eprint = "2009.01693",
    archivePrefix = "arXiv",
    primaryClass = "physics.ins-det",
    reportNumber = "CERN-LHCC-2020-014, LHCC-I-031-ADD-1",
    month = "9",
    year = "2020"
}

@article{Chou:2016lxi,
    author = "Chou, John Paul and Curtin, David and Lubatti, H. J.",
    title = "{New Detectors to Explore the Lifetime Frontier}",
    eprint = "1606.06298",
    archivePrefix = "arXiv",
    primaryClass = "hep-ph",
    doi = "10.1016/j.physletb.2017.01.043",
    journal = "Phys. Lett. B",
    volume = "767",
    pages = "29--36",
    year = "2017"
}

@article{MATHUSLA:2018bqv,
    author = "Alpigiani, Cristiano and others",
    collaboration = "MATHUSLA",
    title = "{A Letter of Intent for MATHUSLA: A Dedicated Displaced Vertex Detector above ATLAS or CMS.}",
    eprint = "1811.00927",
    archivePrefix = "arXiv",
    primaryClass = "physics.ins-det",
    reportNumber = "CERN-LHCC-2018-025, LHCC-I-031",
    month = "7",
    year = "2018"
}

@article{FASER:2022hcn,
    author = "Abreu, Henso and others",
    collaboration = "FASER",
    title = "{The FASER detector}",
    eprint = "2207.11427",
    archivePrefix = "arXiv",
    primaryClass = "physics.ins-det",
    reportNumber = "CERN-FASER-2022-001",
    doi = "10.1088/1748-0221/19/05/P05066",
    journal = "JINST",
    volume = "19",
    number = "05",
    pages = "P05066",
    year = "2024"
}

@article{Salin:2927003,
      author        = "Salin, Olivier and Barr, Alan and McFayden, Josh and Boyd,
                       Jamie and Vranjes Milosavljevic, Marija and Vranjes, Nenad
                       and D'Onofrio, Monica and Gwilliam, Carl and Anders, John
                       Kenneth and Jakobsen, Sune and Otono, Hidetoshi and
                       Queitsch-Maitland, Michaela and Wilson, Benjamin James and
                       Lohwasser, Kristin and Diwan, Milind Vaman and Vicenzi,
                       Matteo and Wu, Wenjie",
      title         = "{FASER2: Detector Design and Performance}",
      year          = "2025",
      url           = "https://cds.cern.ch/record/2927003",
}

@article{FASER:2023tle,
    author = "Abreu, Henso and others",
    collaboration = "FASER",
    title = "{Search for dark photons with the FASER detector at the LHC}",
    eprint = "2308.05587",
    archivePrefix = "arXiv",
    primaryClass = "hep-ex",
    reportNumber = "CERN-EP-2023-161",
    doi = "10.1016/j.physletb.2023.138378",
    journal = "Phys. Lett. B",
    volume = "848",
    pages = "138378",
    year = "2024"
}

@article{FASER:2024bbl,
    author = "Mammen Abraham, Roshan and others",
    collaboration = "FASER",
    title = "{Shining light on the dark sector: search for axion-like particles and other new physics in photonic final states with FASER}",
    eprint = "2410.10363",
    archivePrefix = "arXiv",
    primaryClass = "hep-ex",
    reportNumber = "CERN-EP-2024-262",
    doi = "10.1007/JHEP01(2025)199",
    journal = "JHEP",
    volume = "01",
    pages = "199",
    year = "2025"
}

@article{Hirsch:2020klk,
    author = "Hirsch, Martin and Wang, Zeren Simon",
    title = "{Heavy neutral leptons at ANUBIS}",
    eprint = "2001.04750",
    archivePrefix = "arXiv",
    primaryClass = "hep-ph",
    reportNumber = "APCTP Pre2020-002, IFIC/20-01",
    doi = "10.1103/PhysRevD.101.055034",
    journal = "Phys. Rev. D",
    volume = "101",
    number = "5",
    pages = "055034",
    year = "2020"
}

@article{Wang:2024mrc,
    author = "Wang, Zeren Simon and Zhang, Yu and Liu, Wei",
    title = "{Searching for heavy neutral leptons coupled to axion-like particles at the LHC far detectors and SHiP}",
    eprint = "2409.18424",
    archivePrefix = "arXiv",
    primaryClass = "hep-ph",
    doi = "10.1007/JHEP01(2025)070",
    journal = "JHEP",
    volume = "01",
    pages = "070",
    year = "2025"
}

@article{Gunther:2023vmz,
    author = {G{\"u}nther, Julian Y. and de Vries, Jordy and Dreiner, Herbi K. and Wang, Zeren Simon and Zhou, Guanghui},
    title = "{Long-lived neutral fermions at the DUNE near detector}",
    eprint = "2310.12392",
    archivePrefix = "arXiv",
    primaryClass = "hep-ph",
    doi = "10.1007/JHEP01(2024)108",
    journal = "JHEP",
    volume = "01",
    pages = "108",
    year = "2024"
}

@article{DeVries:2020jbs,
    author = {De Vries, Jordy and Dreiner, Herbert K. and G{\"u}nther, Julian Y. and Wang, Zeren Simon and Zhou, Guanghui},
    title = "{Long-lived Sterile Neutrinos at the LHC in Effective Field Theory}",
    eprint = "2010.07305",
    archivePrefix = "arXiv",
    primaryClass = "hep-ph",
    reportNumber = "APCTP Pre2020-027, BONN-TH-2020-10, RBRC-1328",
    doi = "10.1007/JHEP03(2021)148",
    journal = "JHEP",
    volume = "03",
    pages = "148",
    year = "2021"
}

@article{Beltran:2025oqj,
    author = "Beltr{\'a}n, Rebeca and Hati, Chandan and Hirsch, Martin and Mart{\'\i}n-Gal{\'a}n, Ana",
    title = "{Long-Lived HNLs via ALP Portal at the LHC}",
    eprint = "2510.26946",
    archivePrefix = "arXiv",
    primaryClass = "hep-ph",
    month = "10",
    year = "2025"
}

@unpublished{DDC_github,
    author = "Domingo, Florian and G{\"u}nther, Julian and Kim, Jong Soo and Wang, Zeren Simon",
	month = {December},
	title = {{Displaced Decay Counter: \url{https://github.com/wzeren/Displaced-Decay-Counter}}},
	year = {2025}
}

@unpublished{cern_medium_term_plan,
    author = "CERN",
	month = {December},
	title = {{CERN Medium-Term Plan for the period 2026--2030: \url{https://cds.cern.ch/record/2941417/files/English.pdf}}},
	year = {2025}
}

@article{Sjostrand:2014zea,
    author = {Sj{\"o}strand, Torbj{\"o}rn and Ask, Stefan and Christiansen, Jesper R. and Corke, Richard and Desai, Nishita and Ilten, Philip and Mrenna, Stephen and Prestel, Stefan and Rasmussen, Christine O. and Skands, Peter Z.},
    title = "{An introduction to PYTHIA 8.2}",
    eprint = "1410.3012",
    archivePrefix = "arXiv",
    primaryClass = "hep-ph",
    reportNumber = "LU-TP-14-36, MCNET-14-22, CERN-PH-TH-2014-190, FERMILAB-PUB-14-316-CD, DESY-14-178, SLAC-PUB-16122",
    doi = "10.1016/j.cpc.2015.01.024",
    journal = "Comput. Phys. Commun.",
    volume = "191",
    pages = "159--177",
    year = "2015"
}

@article{ATLAS:2012yve,
    author = "Aad, Georges and others",
    collaboration = "ATLAS",
    title = "{Observation of a new particle in the search for the Standard Model Higgs boson with the ATLAS detector at the LHC}",
    eprint = "1207.7214",
    archivePrefix = "arXiv",
    primaryClass = "hep-ex",
    reportNumber = "CERN-PH-EP-2012-218",
    doi = "10.1016/j.physletb.2012.08.020",
    journal = "Phys. Lett. B",
    volume = "716",
    pages = "1--29",
    year = "2012"
}

@article{CMS:2012qbp,
    author = "Chatrchyan, Serguei and others",
    collaboration = "CMS",
    title = "{Observation of a New Boson at a Mass of 125 GeV with the CMS Experiment at the LHC}",
    eprint = "1207.7235",
    archivePrefix = "arXiv",
    primaryClass = "hep-ex",
    reportNumber = "CMS-HIG-12-028, CERN-PH-EP-2012-220",
    doi = "10.1016/j.physletb.2012.08.021",
    journal = "Phys. Lett. B",
    volume = "716",
    pages = "30--61",
    year = "2012"
}

@article{Nilles:1983ge,
    author = "Nilles, Hans Peter",
    title = "{Supersymmetry, Supergravity and Particle Physics}",
    reportNumber = "UGVA-DPT-1983-12-412",
    doi = "10.1016/0370-1573(84)90008-5",
    journal = "Phys. Rept.",
    volume = "110",
    pages = "1--162",
    year = "1984"
}

@article{Martin:1997ns,
    author = "Martin, Stephen P.",
    editor = "Kane, Gordon L.",
    title = "{A Supersymmetry primer}",
    eprint = "hep-ph/9709356",
    archivePrefix = "arXiv",
    reportNumber = "FERMILAB-PUB-97-425-T",
    doi = "10.1142/9789812839657_0001",
    journal = "Adv. Ser. Direct. High Energy Phys.",
    volume = "18",
    pages = "1--98",
    year = "1998"
}

@inproceedings{Sonneveld:2025ckk,
    author = "Sonneveld, Jory",
    collaboration = "ATLAS, CMS",
    title = "{SUSY Highlights: Current Results and Future Prospects}",
    eprint = "2507.16400",
    archivePrefix = "arXiv",
    primaryClass = "hep-ex",
    month = "7",
    year = "2025"
}

@article{Baer:2025zqt,
    author = "Baer, Howard and Barger, Vernon and Bolich, Jessica and Dutta, Juhi and Martinez, Dakotah and Salam, Shadman and Sengupta, Dibyashree and Zhang, Kairui",
    title = "{Prospects for supersymmetry at High-Luminosity LHC}",
    eprint = "2502.10879",
    archivePrefix = "arXiv",
    primaryClass = "hep-ph",
    reportNumber = "OUHEP-250130",
    doi = "10.1103/bzw1-gfs1",
    journal = "Rev. Mod. Phys.",
    volume = "97",
    number = "4",
    pages = "045001",
    year = "2025"
}

@inbook{Giudice:2017pzm,
    author = "Giudice, Gian Francesco",
    editor = "Levy, Aharon and Forte, Stefano and Ridolfi, Giovanni",
    title = "{The Dawn of the Post-Naturalness Era}",
    booktitle = "{From My Vast Repertoire ...}: {Guido Altarelli's Legacy}",
    eprint = "1710.07663",
    archivePrefix = "arXiv",
    primaryClass = "physics.hist-ph",
    reportNumber = "CERN-TH-2017-205",
    doi = "10.1142/9789813238053_0013",
    pages = "267--292",
    year = "2019"
}

@article{Alimena:2019zri,
    author = "Alimena, Juliette and others",
    title = "{Searching for long-lived particles beyond the Standard Model at the Large Hadron Collider}",
    eprint = "1903.04497",
    archivePrefix = "arXiv",
    primaryClass = "hep-ex",
    doi = "10.1088/1361-6471/ab4574",
    journal = "J. Phys. G",
    volume = "47",
    number = "9",
    pages = "090501",
    year = "2020"
}

@article{Lee:2018pag,
    author = "Lee, Lawrence and Ohm, Christian and Soffer, Abner and Yu, Tien-Tien",
    title = "{Collider Searches for Long-Lived Particles Beyond the Standard Model}",
    eprint = "1810.12602",
    archivePrefix = "arXiv",
    primaryClass = "hep-ph",
    doi = "10.1016/j.ppnp.2019.02.006",
    journal = "Prog. Part. Nucl. Phys.",
    volume = "106",
    pages = "210--255",
    year = "2019",
    note = "[Erratum: Prog.Part.Nucl.Phys. 122, 103912 (2022)]"
}

@article{Beacham:2019nyx,
    author = "Beacham, J. and others",
    title = "{Physics Beyond Colliders at CERN: Beyond the Standard Model Working Group Report}",
    eprint = "1901.09966",
    archivePrefix = "arXiv",
    primaryClass = "hep-ex",
    reportNumber = "CERN-PBC-REPORT-2018-007",
    doi = "10.1088/1361-6471/ab4cd2",
    journal = "J. Phys. G",
    volume = "47",
    number = "1",
    pages = "010501",
    year = "2020"
}

@inproceedings{Kwon:2025fgb,
    author = "Kwon, Hyejin",
    collaboration = "ATLAS, CMS",
    title = "{Supersymmetry searches at ATLAS and CMS}",
    eprint = "2506.06839",
    archivePrefix = "arXiv",
    primaryClass = "hep-ex",
    reportNumber = "CMS-CR-2025-123",
    month = "6",
    year = "2025"
}

@article{ATLAS:2025lfx,
    author = "Aad, Georges and others",
    collaboration = "ATLAS",
    title = "{Search for emerging jets in $pp$ collisions at $\sqrt{s} = 13$ TeV with the ATLAS experiment}",
    eprint = "2510.12347",
    archivePrefix = "arXiv",
    primaryClass = "hep-ex",
    reportNumber = "CERN-EP-2025-225",
    month = "10",
    year = "2025"
}

@article{ATLAS:2025pak,
    author = "Aad, Georges and others",
    collaboration = "ATLAS",
    title = "{Search for events with one displaced vertex from long-lived neutral particles decaying into hadronic jets in the ATLAS muon spectrometer in pp collisions at s=13{\,}{\,}TeV}",
    eprint = "2503.20445",
    archivePrefix = "arXiv",
    primaryClass = "hep-ex",
    reportNumber = "CERN-EP-2025-062",
    doi = "10.1103/cmql-s9sq",
    journal = "Phys. Rev. D",
    volume = "112",
    number = "9",
    pages = "092001",
    year = "2025"
}

@article{CMS:2025urb,
    author = "Chekhovsky, Vladimir and others",
    collaboration = "CMS",
    title = "{Search for vector-like leptons with long-lived particle decays in the CMS muon system in proton-proton collisions at $\sqrt{\text{s}}$ = 13 TeV}",
    eprint = "2503.16699",
    archivePrefix = "arXiv",
    primaryClass = "hep-ex",
    reportNumber = "CMS-EXO-23-015, CERN-EP-2025-021",
    doi = "10.1007/JHEP08(2025)156",
    journal = "JHEP",
    volume = "08",
    pages = "156",
    year = "2025"
}

@article{CMS:2025qkk,
    author = "Hayrapetyan, Aram and others",
    collaboration = "CMS",
    title = "{Search for long-lived particles using displaced vertices with low-momentum tracks in proton-proton collisions at $\sqrt{s}$ = 13 TeV}",
    eprint = "2511.08212",
    archivePrefix = "arXiv",
    primaryClass = "hep-ex",
    reportNumber = "CMS-EXO-24-033, CERN-EP-2025-238",
    month = "11",
    year = "2025"
}

@article{LHCb:2021dyu,
    author = "Aaij, R. and others",
    collaboration = "LHCb",
    title = "{Search for massive long-lived particles decaying semileptonically at ${\sqrt{s}}=13\,\hbox {TeV}$}",
    eprint = "2110.07293",
    archivePrefix = "arXiv",
    primaryClass = "hep-ex",
    reportNumber = "LHCb-PAPER-2021-028, CERN-EP-2021-186",
    doi = "10.1140/epjc/s10052-022-10186-3",
    journal = "Eur. Phys. J. C",
    volume = "82",
    number = "4",
    pages = "373",
    year = "2022"
}

@article{LHCb:2025ymr,
    author = "Aaij, Roel and others",
    collaboration = "LHCb",
    title = "{Search for heavy neutral leptons in B-meson decays}",
    eprint = "2512.14551",
    archivePrefix = "arXiv",
    primaryClass = "hep-ex",
    reportNumber = "LHCb-PAPER-2025-042, CERN-EP-2025-264",
    month = "12",
    year = "2025"
}

@article{Abdullahi:2022jlv,
    author = "Abdullahi, Asli M. and others",
    title = "{The present and future status of heavy neutral leptons}",
    eprint = "2203.08039",
    archivePrefix = "arXiv",
    primaryClass = "hep-ph",
    reportNumber = "FERMILAB-CONF-22-184-T-V",
    doi = "10.1088/1361-6471/ac98f9",
    journal = "J. Phys. G",
    volume = "50",
    number = "2",
    pages = "020501",
    year = "2023"
}

@article{Jeanty:2025wai,
    author = "Jeanty, Laura and Shuve, Brian",
    title = "{Long-lived particles: theory and experimental probes}",
    eprint = "2511.17934",
    archivePrefix = "arXiv",
    primaryClass = "hep-ph",
    month = "11",
    year = "2025"
}

@article{Ema:2025bww,
    author = "Ema, Yohei and Fox, Patrick J. and Hostert, Matheus and Menzo, Tony and Pospelov, Maxim and Ray, Anupam and Zupan, Jure",
    title = "{Long-lived Axion-Like Particles from Tau Decays}",
    eprint = "2507.15271",
    archivePrefix = "arXiv",
    primaryClass = "hep-ph",
    reportNumber = "CERN-TH-2025-123, FERMILAB-PUB-25-0408-T, N3AS-25-011",
    month = "7",
    year = "2025"
}

@article{Patrone:2025fwk,
    author = "Patrone, Samuel and Blinov, Nikita and Plestid, Ryan",
    title = "{Long-lived axion-like particles from electromagnetic cascades}",
    eprint = "2509.14310",
    archivePrefix = "arXiv",
    primaryClass = "hep-ph",
    reportNumber = "CALT-TH/2025-029, CERN-TH-2025-178",
    month = "9",
    year = "2025"
}

@article{Shrock:1980vy,
    author = "Shrock, R. E.",
    title = "{New Tests For, and Bounds On, Neutrino Masses and Lepton Mixing}",
    reportNumber = "ITP-SB-80-23",
    doi = "10.1016/0370-2693(80)90235-X",
    journal = "Phys. Lett. B",
    volume = "96",
    pages = "159--164",
    year = "1980"
}

@article{Shrock:1980ct,
    author = "Shrock, Robert E.",
    title = "{General Theory of Weak Leptonic and Semileptonic Decays. 1. Leptonic Pseudoscalar Meson Decays, with Associated Tests For, and Bounds on, Neutrino Masses and Lepton Mixing}",
    reportNumber = "ITP-SB-80-56",
    doi = "10.1103/PhysRevD.24.1232",
    journal = "Phys. Rev. D",
    volume = "24",
    pages = "1232",
    year = "1981"
}

@article{Shrock:1981wq,
    author = "Shrock, Robert E.",
    title = "{General Theory of Weak Processes Involving Neutrinos. 2. Pure Leptonic Decays}",
    reportNumber = "ITP-SB-81-3",
    doi = "10.1103/PhysRevD.24.1275",
    journal = "Phys. Rev. D",
    volume = "24",
    pages = "1275",
    year = "1981"
}

@article{Bondarenko:2018ptm,
    author = "Bondarenko, Kyrylo and Boyarsky, Alexey and Gorbunov, Dmitry and Ruchayskiy, Oleg",
    title = "{Phenomenology of GeV-scale Heavy Neutral Leptons}",
    eprint = "1805.08567",
    archivePrefix = "arXiv",
    primaryClass = "hep-ph",
    doi = "10.1007/JHEP11(2018)032",
    journal = "JHEP",
    volume = "11",
    pages = "032",
    year = "2018"
}

@article{Mohapatra:1974gc,
    author = "Mohapatra, R. N. and Pati, Jogesh C.",
    title = "{A Natural Left-Right Symmetry}",
    reportNumber = "CCNY-HEP-74-2",
    doi = "10.1103/PhysRevD.11.2558",
    journal = "Phys. Rev. D",
    volume = "11",
    pages = "2558",
    year = "1975"
}

@article{Bando:1998ww,
    author = "Bando, Masako and Yoshioka, Koichi",
    title = "{Sterile neutrinos in a grand unified model}",
    eprint = "hep-ph/9806400",
    archivePrefix = "arXiv",
    reportNumber = "KUNS-1516, HE-TH-98-10",
    doi = "10.1143/PTP.100.1239",
    journal = "Prog. Theor. Phys.",
    volume = "100",
    pages = "1239--1250",
    year = "1998"
}

\end{document}